\documentclass[]{article}
\usepackage{amsmath}
\usepackage{graphicx}
\usepackage{amssymb}
\usepackage[utf8]{inputenc}

\usepackage{tgbonum}
\newcommand{\Ai}{{\mathrm{Ai}}}
\newcommand{\Bi}{{\mathrm{Bi}}}
\newcommand{\e}{{\mathrm{e}}}
\numberwithin{equation}{section}
\everymath{\displaystyle}
\title{Tomographic analysis of the De Sitter model in quantum and classical cosmology}
\author{C. Stornaiolo\\
{\em $~^{3}$ Istituto Nazionale di Fisica Nucleare,}\\
{\em Sezione di Napoli,}\\
 {\em  Complesso Universitario di Monte S. Angelo}\\
 {Edificio 6, via Cinthia,45 -- 80126 Napoli, Italy} \\ }

\begin{document}

\maketitle

\begin{abstract}

The importance of the tomographic approach is that either in quantum mechanics as in classical mechanics the state of a physical system is expressed with the same family of functions, the tomograms.
The extension of this procedure to quantum cosmology is straightforward. But instead of using the tomographic representation, we use tomograms to analyze the properties of the quantum and classical universes, starting from the wave functions in quantum cosmology and the phase space distribution in classical cosmology.
In this paper we resume the properties of the tomographic approach  introduced in previous papers.
 Then we  study and discuss the properties of the initial conditions introduced by Hartle and Hawking and by Vilenkin and Linde  and we   study their classical transition. It results that a possible reason for the quantum to classical transition is the decay of the cosmological constant from the Planck scale to the present one. So that  the Cosmological Constant Problem becomes a crucial topic  in study of the evolution of the universe.
 
\end{abstract}

\section{Introduction}
General Relativity is a well-established theory which introduces a close relation between geometry and physics. However its undeniable successes are in some cases counterbalanced by some of its shortcomings.

The  Hawking and Penrose  theorems show that under conditions of the energy-momentum tensor, which are compatible with alle the forms of matter fields we know, the solutions of the Einstein equations must contain at least one singularity. This means that the theory is not always able to make physical predictions because the presence a singularity means that the state of a gravitational system can  not determined. Because as the state of a  physical system is represented by a set of data that allow us  to predict its evolution.   The presence of a singularity  means the absence of such data and therefore the impossibility to predict the physical evolution of the system.

In cosmology the singularity is related to the degeneracy of the metric and of  the scalar curvature and it corresponds to an infinite mass-energy density. This suggests  that at states close to the singularity  General Relativity  can not be a  reliable theory.
   
The high densities and small distances involved suggest that it is most likely to be a quantum theory that has general relativity as its classical limit.
   
In the quest of this theory we can apply the canonical quantization of  the gravitational field, according to the prescriptions introduced by Dirac in 1932.   This program was completed by Arnowitt, Deser and Misner\cite{Arnowitt:1962hi}  by introducing the Hamitonian formulation of general relativity  where    the spatial components of the  metric and their respective conjugate momenta defined on a spatial 3-surface $ S $ are the canonical variables.  The Einstein equations are equivalent to two   classes of equations, the evolution equation obtained the Hamiltonian equation of motion, but  the properties of the theory imply too the existence of  constraint equations among the variables which, according to the Bianchi identities,  are preserved during the evolution of the system on a  each spatial 3-surface $ S $.  

In quantum gravity the role of the Schroedinger equations is taken by the quantum constraint equations (the Wheeler De Witt equation and the diffoemorphism equation), which become  operators that annihilate the wave functionals. A direct development of this formalismo is loop quantum gravity that will be discussed elsewhere.

 A relevant chapter of quantum gravity is quantum cosmology. It is considered as a simplified version of  quantum gravity because it takes as canonical variables  only homogeneous spatial metrics and there conjugate momenta. This restriction reduces the infinite degrees of freedom of the quantum gravity  theory  to a finite number,  by this simplification  many properties of the general theory can be explored with more ease. But the study of quantum cosmology has become an important issue in modern physics because  it can explain the origin of the large structure of the universe.
 
The two fundamental laws at the basis of  quantum cosmology theory  are  the dynamical description of the quantum evolution of the universe and the assignment of the  initial quantum state. 
 
 The Wheeler-DeWitt equation equation  in quantum cosmology plays a role   similar to  the Schr\"{o}dinger equation in quantum mechanics. 
 
Among the  infinite  solutions of the Wheeler-De Witt equation the physical  ones  are those which have a classical limit  and predict the large scale of the universe observed. . According to our present knowledge  we live in a almost homogeneouos and isotropic universe, very likely flat, whose particular  structure was originated by  primordial quantum  perturbations generated during the inflationary epoch. The acceleration of the observed expansion recently completes the current picture of the state of the universe. The origin of this expansion has not yet been definitively explained and is presumably attributable either to a form of dark energy or to a substantial modification of Einstein's equations. 

This scenario must be the result of the evolution of a classical primordial universe descending from a quantum universe whose properties are in turn determined  by particular initial conditions.  Such initial conditions must be regarded  as a fundamental physical law, because unlike an ordinary physical system they can not be arbitrarily assigned in relation to the conditions designed by an observer. They must select the only quantum state that conducts to the correct evolution of the universe.

The choice of the initial state of the universe can be deduced from particular physical requirements, first imposing  the absence of an initial singularity and second  by requiring that  the  collapsing modes will not make the newborn universe recollapses. Alternatively one  can also think that some informations of the initial stages of the universe survive on the present structure and that they can be iderived  from observational data.

 In this article we will discuss the  initial conditions proposed by Hartle and Hawking, Vilenkin and Linde, in particular we consider them, for sake of simplicity, in the context of a  de Sitter, i. e.  just in  presence of a cosmological constant. We  analyze the different solutions in the context of quantum tomography. It is interesting to note that greater versatility is obtained in the understanding of an argument, when the different representations of quantum mechanics are used simultaneously.

Quantum tomography was  introduced in very recent times  to have a means to reconstruct the initial state of a quantum system by detecting  probability  distribution functions. An observer gets a statistical prediction of the values of any observable as the result of the measurements within the limits imposed by quantum mechanics. 

To this purpose q uantum tomography has been applied  quantum optics by  using the properties of balanced homodyne detection\cite{Vogel:1989zz} \cite{bertrand} \cite{breitenbach97}.
The homodyne measurement of an electromagnetic field gives all the possible linear combinations of the field quadratures just by varying the phase of a local oscillator.
At any phase of the local oscillator the average of the outcomes is related to  a marginal distribution of the Wigner function \cite{mancini95}.
This is the most straightforward way  to reconstruct experimentally the Wigner function and consequently  any other quasi probabilistic distribution.

Similarly quantum tomography is applied to quantum mechanics  to mount an atomic optics apparatus to reconstruct the initial state of ta quantum system (see \cite{Janicke:1989}  for the problem in one dimension and  \cite{Raymer:1994} for the problem   in two dimensions) . The experiment was realized by \cite{Kurtsiefer:1997}  in  reconstructing the initial state of a helium atom(see also \cite{Kurtsiefee:1997a} for the realization of the detector).

Theoretical research has shown that conventional quantum mechanics can have a representation in which the fundamental quantities describing the quantum states are the tomograms instead of the wave function. In many respects it is  useful  the tomographic representation cf quantum mechanics, because tomograms have an immediate physical interpretation because they are marginal probability  functions therefore they are observables. The other important aspect is that  one can define  classical tomograms on the phase space and we can describe the physical state of a system in a single way both in quantum mechanics and in classical mechanics.

We can to extend quantum tomography to cosmology. It seems an appropriate way to address the issue of the initial conditions problem. We will show how to construct quantum and classical tomograms, study their properties and then consider the transition from quantum to classical, in order to try to connect  the properties of the large scale universe with the initial conditions.

In sect. \ref{introducttomography} symplectic tomography is introduced. We introduce the definition of a quantum tomogram as the Radon transform of the Wigner function. From this we find also the relation of tomogram and wave function. In the following are listed all the properties of the tomograms and  how to calculate them  with the stationary phaase approximation.  The definition of classical tomograms ends the section.

Classical cosmology is introduced in sect. \ref{desitteruniverse}. After a very short resume of the relevant properties of the de Sitter universe, we derive the tomogram after defining the probability distribution on the phase space through the Hamiltonian classical contraint.  
In sect.\ref{desitterquantum} we consider the wave equations corresponding to the  proposals of Hartle-Hawking, Vilennkin and Linde as  solutions of the Airy equations.
But only in the first case it is possible to express the Wigner function as an Airy function as well. The other two finctions are not limited so it is not possible to perform an integral transform. The same is happens in the definition of the tomograms.We conclude the section by taking the classical limit.The arguments developed in this paper open new perspectives in the  study of quantum cosmology. In sect \ref{conclusions}  after resuming the results of this paper, we list the possible developments of this work, which was conceived originally to relate the data on the large scale structure of the universe with the initial conditions, but it shows also the  power of analysis of tomography in understanding 	quantum cosmology. 

\section{Introduction to symplectic tomography}\label{introducttomography}
 In classical mechanics the state of a physical system is given by that set of data that allow us to predict the evolution of this system on the basis of the physical laws that rule them.
 
  For example, the initial position and velocity of a point particle in a potential allows to determine its the motion, which in turn is described as a succession of states determined uniquely by the equation of motion.
  
  In phase space   the state of a  single classical point particle is given by the distribution function
  \begin{equation}\label{distriburionparticle1}
  f(q,p)=\delta(q-q_{0})\delta(p-p_{0}).
  \end{equation}  
 
 Due to the uncertainty principle according to which  it is not possible to fix with arbitrarily high precision the position and the momentum of a particle in the same time,  this definition of physical state does not make sense in quantum mechanics. Rather  the state of a system is intended in terms of  a probability distribution for each observable.  
 
 According to \cite{nine} there are nine formulations of quantum mechanics, the tomographic representation is the tenth. In each  of these formulations the physical state of a quantum system is represented differently.
    

For example   Wigner introduced a function on phase space defined by the   transform of the wave function  $ \psi(x) $ 
\begin{equation}\label{Wigner function}
W(x,p)=\frac{1}{2\pi}\int \psi\left(x+\frac{\hbar u}{2}\right) \psi^{*}\left( x-\frac{\hbar u}{2}\right)\e^{-iup} du \, .
\end{equation}
to  describe the state of a  quantum particle on the phase space in analogy with classical mechanics. The Wigner function is defined with the requirement that it has to be normalized

\begin{equation}\label{normalwignerfunction}
\int W(q,p)dq\, dp=1.
\end{equation}

In quantum cosmology alongside the concept of wave function of the universe, the density matrix representation and the  phase space representations have been introduced with the purpose  of  addressing the problem of the classical  transition  for a quantum universe.  
 
Unfortunately, these formalisms can  hardly be  directly related to the observational data, because the wave function and the density matrix do not have a   physical interpretation, while  the Wigner function cannot  be considered  a probability function as it can take negative values unlike  the Boltzmann which instead  is probability  distribution function on phase space,  

Recently the notion of standard positive probability distribution function (tomogram) was introduced in cosmology to describe the quantum state of universe as an  alternative to wave function or to density matrix .

  The tomogram  can be defined by projecting   the Wigner function on the  axes on the phase space at arbitrary angles. Mathematically this is represented by the modified Radon transform,
 \begin{equation}\label{tomowigner}
 	\mathcal{W}\left(X,\mu,\nu \right)=
  \int W(q,p)\delta(X-\mu x-\nu p)dx dp\,.,
 \end{equation}
 
 The function $	\mathcal{W}\left(X,\mu,\nu \right) $ depends on the random variable $ X $ and two real parameters $ \mu$ and $\nu  $.  Specifically $ X $  is given by a linear combination of position and momentum:
\begin{equation}\label{X}
X=\mu x + \nu p
\end{equation}  
with $ \mu=s \cos \theta $ and $ \nu=s^{-1} \sin \theta $.
 
 So, we say that the quantum state is given if the position probability distribution $ \mathcal{W} ( X, \mu, \nu) $ in an ensemble of rotated and squeezed reference frames in classical phase space is given. Given a large ensemble of particles prepared in the same state, it gives the probability that the pseudo classical trajectory  of a single particle depends on a pair of initial conditions contained in the combination (\ref{X}) .
 
 From (\ref{tomowigner}) we can  derive the relation between a wave function $ \psi(x) $ and  its  corresponding tomogram. We must consider separately the case $ \nu\ne 0 $ and  the case $ \nu=0 $.
 
Let us consider first $ \nu\neq 0 $.  It is convenient to write the Wigner function  in the following form
\begin{equation}\label{key}
 	W(x,p)=\frac{1}{2\pi}\int \psi\left( y'\right) \psi^{*}\left( y \right) \e^{ip(y'-y)}d(y'-y)
 \end{equation}
 so that the tomogram is written as 
 
\begin{equation}\label{tomowigner}
 	\mathcal{W}\left(X,\mu,\nu \right)=
 	\frac{1}{2\pi}\int \psi\left( y\right) \psi^{*}\left( y' \right) \e^{ip(y'-y)}\delta(X-\mu x-\nu p)d(y'-y)dx dp
 \end{equation}
  
where
\begin{align}
y=x-\frac{u}{2}&\ \ \ \ \ \ \ \  x=\frac{y+y'}{2}\\
y'=x+\frac{u}{2}& \ \ \ \ \ \ \ \ u=y'-y.
\end{align}
Applying the definition of tomogram we find
 \begin{align} \label{nunonzero}
\nonumber	\mathcal{W}\left(X,\mu,\nu \right)&=
 	\frac{1}{2\pi}\int \psi\left( y'\right) \psi^{*}\left( y \right) \e^{ip(y'-y)}\delta(X-\mu x-\nu p)d(y'-y)dx dp\\ 
    \nonumber	&=\frac{1}{2\pi|\nu|}\int \psi\left( y\right) \psi^{*}\left( y' \right) \e^{i\left( \frac{X}{\nu}-\frac{\mu}{2\nu}(y+y')\right) (y'-y)}
 	d(y'-y)dx \\  \,
 	  \nonumber &=\frac{1}{2\pi|\nu|}\int \psi\left( y\right) \psi^{*}\left( y' \right) \e^{i\left( \frac{X}{\nu}(y-y')-\frac{\mu}{2\nu}(y^{2}-y'^{2})\right)}
 	dy'dy	\\
 	 & \equiv\frac{1}{2\pi|\nu|}\left| \int \psi\left( y\right)  \exp\left[ {i\left( \frac{\mu}{2\nu}y^{2}-\frac{X}{\nu}y\right)}\right] \right| ^{2}
 \end{align}

 If $ \nu=0 $ we have
 \begin{align}
 \nonumber	\mathcal{W}\left(X,\mu,0 \right)
	&=\frac{1}{2\pi}\int \psi\left( x+\frac{u}{2}\right) \psi^{*}\left( x-\frac{u}{2}\right) \e^{iup}\delta(X-\mu x)dx dp du\\
\nonumber 	&=\frac{1}{2\pi}\int \psi\left( x+\frac{u}{2}\right) \psi^{*}\left( x-\frac{u}{2}\right) \delta(u)\delta(X-\mu x)dx du\\
 \nonumber	&=\frac{1}{2\pi}\int \psi\left( x\right) \psi^{*}\left( x\right) \delta(X-\mu x)dx\\
 	&=\frac{1}{2\pi|\mu|}\left|  \psi\left( \frac{X}{\mu}\right)  \right| ^{2}= \frac{1}{2\pi|\mu|}\left|  \psi\left( x\right)  \right| ^{2},
 \end{align}
 because with $ \nu=0 $
 \begin{equation}\label{key}
 X=\mu X
 \end{equation}
in particular $$ 	\mathcal{W}\left(X,1,0 \right)=\frac{1}{2\pi}\left|  \psi\left(X\right)  \right| ^{2}\equiv \frac{1}{2\pi}\left|  \psi\left(x\right)  \right| ^{2} .$$

It must be noted that the general expression for the tomogram (\ref{nunonzero}) is the square modulus of   the tomographic amplitude
\begin{equation}\label{amplitude}
\mathcal{A}\left(X,\mu,\nu \right)=
\int\psi(y)\exp\left(i\frac{\mu y^{2}}{2\nu}-i\frac{Xy}{\nu} \right) dy  \, .
\end{equation}
which  is the fractional Fourier transform of the wave function.

When $ \mu=0 $ and $ \nu=1 $ the tomographic amplitude reduces to a Fourier transform from $ x $ to $ X\equiv p $, leading   in (\ref{nunonzero})  to the 
density $ |\psi(p)|^{2} $ in the $ p  $ representation. 

The relation  $ X=\mu x + \nu p$  between $ x $, $ p $ and $ X $ is  a collection of  linear canonical transformations parameterized by $ \mu $ and $ \nu $ with generating function
\begin{equation}\label{genfunct}
G_{(\mu\nu)}(x,X)=-\frac{\mu}{2\nu}\left( x-\frac{X}{\mu}\right) ^{2}=-\frac{\mu}{2\nu}x^2+\frac{Xx}{\nu}-\frac{X^{2}}{2\mu\nu} 
\end{equation}
so that (\ref{nunonzero}) can be written as 
\begin{equation}\label{canonicaltomogram}
	\mathcal{W}\left(X,\mu,\nu \right)=\frac{1}{2\pi|\nu|}\left| \int \psi\left( y\right)  \exp\left[ -{i G_{(\mu\nu)}(y,X)}\right]dy \right| ^{2}\, .
\end{equation}
Eq. (\ref{canonicaltomogram} ) can be interpreted as the square modulus of   the wave function obtained by a linear canonical transform  in the new variable  $ X $.

Finally we observe that the relation found between tomogram and wave function  can be applied to  the density matrix of a pure state and extended to any density matrix  in the following way,
\begin{equation}\label{kdensity}
\mathcal{W}\left(X,\mu,\nu \right)
=\frac{1}{2\pi|\nu|} \int\rho(y,y')exp\left[-i\frac{y-y'}{\nu}\left( X-\mu\frac{y+y'}{2}\right)  \right] dydy'\, .
\end{equation}

%
%
%
%

 These relations can all be inverted to reconstruct the Wigner function, the density matrix and the wave function once the tomograms have been determined. As a matter of fact  it can be proved that there exist a one to one correspondence among all these representations of a quantum state.
 
\subsection{Reconstruction of the wave function} 
\textit{Reconstruction of $ \psi(x)  $ from the Wigner function}

\begin{equation}\label{key}
\int W(x,p)\e^{-2ip\xi'/\hbar}dp=\int\psi^{*}(x+\xi)\delta(\xi-\xi')\psi(x-\xi)d\xi=\psi^{*}(x+\xi')\psi(x-\xi')
\end{equation} 
  Let us choose an $ a $ such that
 \begin{equation}\label{key}
 \int W(a,p)dp= \psi^{*}(a)\psi(a')\neq0
 \end{equation}
 and $ \xi'=a-x $,
 
 \begin{equation}\label{key}
 \int W(x,p)\e^{-2ip(a-x)/\hbar} dp=\psi^{*}(a)\psi(2x-a)
 \end{equation}
 
 \[ 2x-a\to x \]
 
 \begin{equation}\label{key}
 \psi(x)=\frac{1}{\psi^{*}(0)}\int W\left( \frac{x+a}{2},p\right) \e^{ip(x-a)/\hbar} dp
 \end{equation}
 
 \begin{equation}\label{key}
 \psi(x)=\frac{1}{\psi^{*}(0)}\int W\left( \frac{x}{2},p\right) \e^{ipx/\hbar} dp
 \end{equation}

 \textit{Reconstruction of the Wigner function from the tomogram}
 
 \begin{equation}\label{key}
 W(x,p)=\int\mathcal{W}(X,\mu,\nu)\e^{i(X-\mu x- \nu p)}dXd\mu d\nu
 \end{equation}
 consequently the recontruction of the wave function from the tomogram descends from the preceding equations, in the case $ a=0 $,
 \begin{equation}\label{key}
 \psi(x)=\frac{\hbar }{2\pi}\frac{1}{\psi^{*}(0)}\int \mathcal{W}(X,\mu,\nu)\e^{i(X-\mu \frac{x}{2}- \nu p+\frac{px}{\hbar})}dXd\mu d\nu dp
 \end{equation}
 Integrating with respect to $ p $ and $\nu  $ we finally obtain
  \begin{equation}\label{key}
  \psi(x)=\frac{\hbar }{2\pi}\frac{1}{\psi^{*}(0)}\int \mathcal{W}(X,\mu,\nu)\e^{i(X-\frac{\mu  x}{2}) )}dXd\mu  \, .
  \end{equation}
  
  \subsection{Properties of the tomograms}
  
  The symplectic tomogram $\mathcal{ W}(X,\mu, \nu)  $ has the properties which
follow from its definition by using the known properties of delta function,
namely,

\noindent
1) Nonnegativity

$$ \mathcal{W}(X,\mu,\nu)\geq 0$$
(this holds by observing that the trace of the product of two
positive operators is a positive number).

\noindent
2) Normalization

\begin{equation}\label{normalization}
	 \int\mathcal{ W}(X,\mu, \nu) dX = 1. 
\end{equation}
These first two  conditions  are important because the tomogram is a probability function.

\noindent
3)  homogeneity

\begin{equation}\label{homogeneity}
\mathcal{ W}(\lambda X,\lambda\mu, \lambda\nu) = \frac{1}{|\lambda|}
\mathcal{ W}(X,\mu, \nu) . 
\end{equation}
  \subsection{Stationary phase approximation for tomograms}\label{stationaryphase}
 If we write the wave function in  the quasi classical approximation, which corresponds to a stationary solution where the energy $ E $ is a constant and the amplitude and the phase $ S $ are connected. $ S $ is a solution of the Hamilton-Jacobi equation. 
  
  \begin{equation}\label{phaseandmodulus}
  \psi(x)=|\psi(x)|\e^{iS(x)}
  \end{equation}
 in $ |\psi(x)| $ varies slowly  the tomogram is given in the following way,  
 \begin{equation}\label{inttomophasemod}
 \mathcal{W}(X,\mu,\nu)=\frac{1}{2\pi|\nu|}\left| \int |\psi\left( y\right)|  \exp\left[ {i\left(S(y)+ \frac{\mu}{2\nu}y^{2}-\frac{X}{\nu}y\right)}\right] \right| ^{2}
 \end{equation}
 and applying the stationary phase approximation to obtain
 \begin{equation}\label{kstationaryphaseapproximation}
 \mathcal{W}(X,\mu,\nu)\approx\frac{1}{2\pi|\nu|}\left| \psi(x_{0})\right|^{2}\left|\left.\frac{\partial^{2}S}{\partial x^{2}} \right| _{x=x_{0}} +\frac{\mu}{\nu}\right| ^{-1} 
 \end{equation}
 where $ x_{0} $ is the solution of equation
 \begin{equation}\label{stationary point x}
 \frac{\partial S}{\partial x} +\frac{\mu}{\nu}x-\frac{X}{\nu}=0.
 \end{equation}
 
 If eq. (\ref{stationary point x}) has multiple roots
  $( x_{0}^{(1)}\dots x_{0}^{(N)} )$ then the tomogram becomes
 
 \begin{equation}\label{kstationaryphaseapproximation}
 \mathcal{W}(X,\mu,\nu)\approx\frac{1}{2\pi|\nu|}\left|\sum_{i=1}^{N} \frac{\psi(x_{0}^{(i)})}{\sqrt{\left. \frac{\partial^{2}S}{\partial x^{2}}\right| _{x=x_{0}^{(i)}}+\frac{\mu}{\nu} }}\right| ^{2} \,.
 \end{equation}
 We notice that  
 \begin{equation}\label{momentum}
 \frac{\mu}{\nu}x-\frac{X}{\nu}=-p
 \end{equation}
 so eq.(\ref{stationary point x})  expresses the condition of classical correlation, 
  \begin{equation}\label{stationary point x}
 p= \frac{\partial S}{\partial x}\, .
  \end{equation}
  The stationary
points, solutions of eq. (\ref{stationary point x}) are they are the points where the integrand is maximum giving the
most important contribution to the value of the integral.  This means that the meaningful  points which lead to the semiclassical solutions are at the peaks of the tomographic amplitudes, which lead to the solutions given in eq.  (\ref{kstationaryphaseapproximation}).
 \subsection{Classical tomograms}
 The definition of tomogram (\ref{tomowigner}) can be extended to any distribution on the phase space. Therefore we can extend this definition to classical distributions $ f(q,p) $, e.g. solutions of the Boltzmann equation,  
 
 \begin{equation}\label{tomoboltz}
  	\mathcal{W}\left(X,\mu,\nu \right)=
   \int f(x,p)\delta(X-\mu x-\nu p)dx dp\,.
  \end{equation}
 If the classical probability distribution $ f(q,p) $ is normalized then also the tomogram is normalizedand satisfies the following conditions,
 \begin{equation}\label{normalization condition tomogram}
 \int \mathcal{W}\left(X,\mu,\nu \right) dX=1
 \end{equation}
 \begin{equation}\label{tomogramposition}
 \mathcal{W}\left(X,1,0 \right) =\int f(x,p)dp
 \end{equation}
 and
 \begin{equation}\label{tomogrammomentum}
 \mathcal{W}\left(X,0,1 \right)=\int f(x,p) dx\,.
 \end{equation}
For example taking the distribution function 
 \begin{equation}\label{classicalparticlestate}
 f(q,p)=\delta(x-x_{0})\delta(p-p_{0})
 \end{equation}
introduced at the beginning of this section, we obtain
 by  eq, (\ref{tomoboltz}) that the tomogram of a particle is
\begin{equation}\label{key}
  \mathcal{W}\left(X,\mu,\nu \right)=\delta(X-\mu x_{0}-\nu p_{0})
\end{equation}
For a moving particle the   time dependent distribution function is
 \begin{equation}\label{key}
 f(q,p,t)=\delta(x-x(t)\delta(p-p(t))
 \end{equation}
 and the corresponding tomogram is 
 \begin{equation}\label{key}
   \mathcal{W}\left(X,\mu,\nu \right)=\delta(X-\mu x(t)-\nu p(t)\,. 
 \end{equation}
 In particular for a free particle with mass $ m $ and given the  equations of motion 
 \begin{align} 
 x(t)&=x_{0}+\frac{p}{m}t\\
 p(t)&=p_{0}= const.
 \end{align}
the particle describes a trajectory on the phase space given by the distribution function
\begin{equation}\label{key}
 f(q,p,t)=\delta\left( x-x_{0}+\frac{p}{m}t\right) \delta(p-p_{0})
 \end{equation}
 and the corresponding  tomogram  is
 \begin{equation}\label{key}
   \mathcal{W}\left(X,\mu,\nu \right)=\delta\left( X-\mu \left( x_{0}+\frac{p_{0}}{m}t\right) -\nu p_{0}\right) \,. 
 \end{equation}
 
 \section{The De Sitter universe}\label{desitteruniverse}
 
The de Sitter model was of crucial importance because it showed that the cosmological constant, contrary to what Einstein believed initially, makes the universe expand even in the absence of material fields. It is given by the equation

\begin{equation}\label{Einstein equation}
G_{\mu\nu}=\Lambda g_{\mu\nu}
\end{equation}
abd the metric is the homogeneous and isotropic metric 
\begin{equation}\label{metric}
ds^{2}=N^{2}c^{2}dt^{2}-a^{2}(t)\left(\frac{dr^{2}}{1-k^{2}r^{2}}+r^{2}(d\theta^{2}+\sin ^{2}\theta d\phi^{2}) \right) 
\end{equation}
where $ N  $ is the lapse, the function that encodes the choice of the time coordinate.
If $ k=0 $ the classical solution gives an exponential expansion 
\begin{equation}\label{exponential expansion}
a(t)\propto e^{Ht}
\end{equation} 
where the Hubble constant $ H $ is
\begin{equation}\label{hubble}
H=\sqrt{\frac{\Lambda}{3}}.
\end{equation}
It must be noted that there is a singularity at $ t= -\infty $, i.e. the cosmological constant violates the Hawking-Penrose conditions. 

The importance of the de Sitter model    in cosmology   was that it confirmed that General Relativity predicted a non static universe. This model was reintroduced with  the inflationary paradigm   to overcome  the  drawbacks  of  traditional cosmological models  which were unable to explain in a natural way the flatness and the homogeneity of the universe. 

Let us consider a closed homogeneous and isotropic  model with a cosmological constant $ \Lambda $  was used in quantum cosmology. Using  the notations  of  \cite{Habib:1990hz}\cite{Habib:1990hx}  \cite{Cordero:2011xa},   we write  the   metric  
\begin{equation}\label{corderodesitter}
ds^{2}= l_{p}^{2}[-N^{2}(t)dt^{2}+a^{2}(t)d\Omega_{3}^{2}]
\end{equation}
where $ a(t) $ is the expansion factor of the universe,  $ d\Omega_{3}^{2} $ is the metric of the unit  three-sphere $ N(t) $ is the lapse function, $l_{p} =2/3 \ell_{p} $, with $ \ell_{p} $ the Planck length. 
Introducing  the following variables
\begin{equation}\label{key}
q=a^{2}\ \ \ \ \ \  \tilde{N}=Nq
\end{equation}
 the Hamiltonian takes the form
\begin{equation}\label{HamiltoniandeSitter}
\mathcal{H}=\frac{1}{2}\left( -4p^{2}+\lambda q-1\right) 
\end{equation}
where $ \lambda $ is the cosmological constant in Planck units.
The ADM formalism implies  the constraint 
\begin{equation}\label{Hamiltonianconstraint}
	\mathcal{H}=0
\end{equation}
this corresponds to  a trajectory on the phase space. At a fixed time the distribution on the phase space reduces to just  one point of this trajectory. 
Differently from the example given in the previous section (see eq (\ref{classicalparticlestate})), position and momentum are not independent. 

The equation of motion tells us how "fast" thie universe runs along this trajectory.

 We represent this trajectory  with  the   distribution
\begin{equation}\label{classicaluniverseconstraint}
f(q,p)=\delta\left( -4p^{2}+\lambda q-1\right) 
\end{equation}
and the corresponding classical  tomogram is
\begin{align}\label{classicaluniversedetomogram}
 \nonumber		\mathcal{W}\left(X,\mu,\nu \right)&=
	\int\delta\left( -4p^{2}+\lambda q-1\right)\delta(X-\mu q-\nu p)dq dp\\
	 &=\frac{1}{2|\mu|}\frac{1}{\left| \sqrt{\frac{\lambda^{2}\nu^{2}}{16\mu^{2}}+\frac{\lambda X}{\mu}-1}\right| }\, .
\end{align}

This tomogram cannot be normalized as it it not integrable on  $ -\infty $ to $ +\infty $. It can be normalized only if  the function (\ref{classicaluniversedetomogram}) has a compact support $ C $.  We  the inferior value of the interval is giving by requiring the reality of the equare root, 
\begin{equation}
\frac{\lambda^{2}\nu^{2}}{16\mu^{2}}+ \frac{\lambda X}{\mu}-1\ge 0
\end{equation}
i.e.
\begin{equation}\label{limite inferiore X}
X\ge \frac{\mu}{\lambda} \left( 1-\frac{\lambda^{2}\nu^{2}}{16\mu^{2}}\right) 
\end{equation}

The superior iimit of $ C $ follows by imposing the normalization of the tomogram (\ref{normalization condition tomogram}).
\begin{equation}
\int_{-\infty}^{\infty} \mathcal{W}(X,\mu,\nu)= \frac{1}{2|\mu|}\int_{\frac{\mu}{\lambda} \left( 1-\frac{\lambda^{2}\nu^{2}}{16\mu^{2}}\right) }^{\lambda\mu+\frac{\mu}{\lambda} \left( 1-\frac{\lambda^{2}\nu^{2}}{16\mu^{2}}\right) }\frac{1}{\left| \sqrt{\frac{\lambda^{2}\nu^{2}}{16\mu^{2}}+\frac{\lambda X}{\mu}-1}\right| }=\frac{\mu}{\left|\mu \right| }\, .
\end{equation}
which is equal to $ 1 $ only for positive $ \mu $
In conclusion we redefine the  function  (\ref{classicaluniversedetomogram})
with  support on the compact interval,
\begin{equation}\label{compact}
 \left[1-\frac{\lambda^{2}\nu^{2}}{16\mu^{2}}, \, \lambda\mu +1-\frac{\lambda^{2}\nu^{2}}{16\mu^{2}}\right] \, .
\end{equation}
as 
\begin{equation}\label{tomogramcompactsupport}
\mathcal{W}\left(X,\mu,\nu \right)=
 	 \begin{cases} 
 	 0 & X < 1-\frac{\lambda^{2}\nu^{2}}{16\mu^{2}} \\
 	 \\
	\frac{1}{2|\mu|}\frac{1}{\left| \sqrt{\frac{\lambda^{2}\nu^{2}}{16\mu^{2}}+\frac{\lambda X}{\mu}-1}\right| } &  1-\frac{\lambda\mu-\nu^{2}}{16\mu^{2}}\le X\le   \lambda\mu +1-\frac{\lambda^{2}\nu^{2}}{16\mu^{2}}  \\
 	 	\\
 	 0 & X>  \lambda\mu +1-\frac{\lambda^{2}\nu^{2}}{16\mu^{2}}
 	 \end{cases}	
 	\end{equation}
 	with $ \mu>0 $.

Let us assume that the cosmological constant has the value currently accepted. In Planck units  $  \lambda=5.6 \cdot 10^{-122}\, t_{p}^{-2} $,  therefore to all practical effects we can consider the function (\ref{tomogramcompactsupport}) as a   Dirac delta. We can show this by considering for sake of simplicity the integral  of any function $ f(x) $ with the function $ 1/(2c \sqrt{x-x_{0}}) $
with support on the closed interval $ [x_{0}, x_{0}+c^{2}] $
\begin{align} \label{teoremino}
\nonumber \frac{1}{2c}\int_{x_{0}}^{x_{0}+c^{2}}\frac{f(x)}{\sqrt{cx-x_{0}}}dx&=\frac{1}{2c}\int_{x_{0}}^{x_{0}+c^{2}}\left( \frac{f(x_{0})}{\sqrt{x-x_{0}}}  + \frac{f'(x_{0})}{\sqrt{x-x_{0}}}  (x-x_{0})+\dots\right) dx\\
\nonumber  &=\frac{1}{2c}\left( 2f(x_{0})\sqrt{x-x_{0}} |_{x_{0}}^{x_{0}+c^{2}} + f'(x_{0})(x-x_{0})^{3/2}|_{x_{0}}^{x_{0}+c^{2}}+\dots\right)\\
&=f(x_{0}) + f'(x_{0}) \times c^{2} + O(c^{4})
\end{align}
if $ c $ is negligible  integration (\ref{teoremino})  gives approximately $ f(x_{0}) $.

In conclusion, if the cosmological constant is of the order of magnitude of the value accepted nowadays  the classical tomogram of a de Sitter  universe  may  be expressed by a  delta function,

\begin{equation}
\mathcal{W}\left( X,\mu,\nu\right) = \delta\left( \frac{\lambda}{\mu}X +\frac{\lambda^{2}\nu^{2}}{16 \mu^{2}}-1\right) \, .
\end{equation}

It is interesting to note that  in the classical  universe the sharp conditions of sharp  homogeneity  are determined by the extreme smallness of the cosmological constant.

\begin{figure}
\centering
\includegraphics[width=1.2\linewidth]{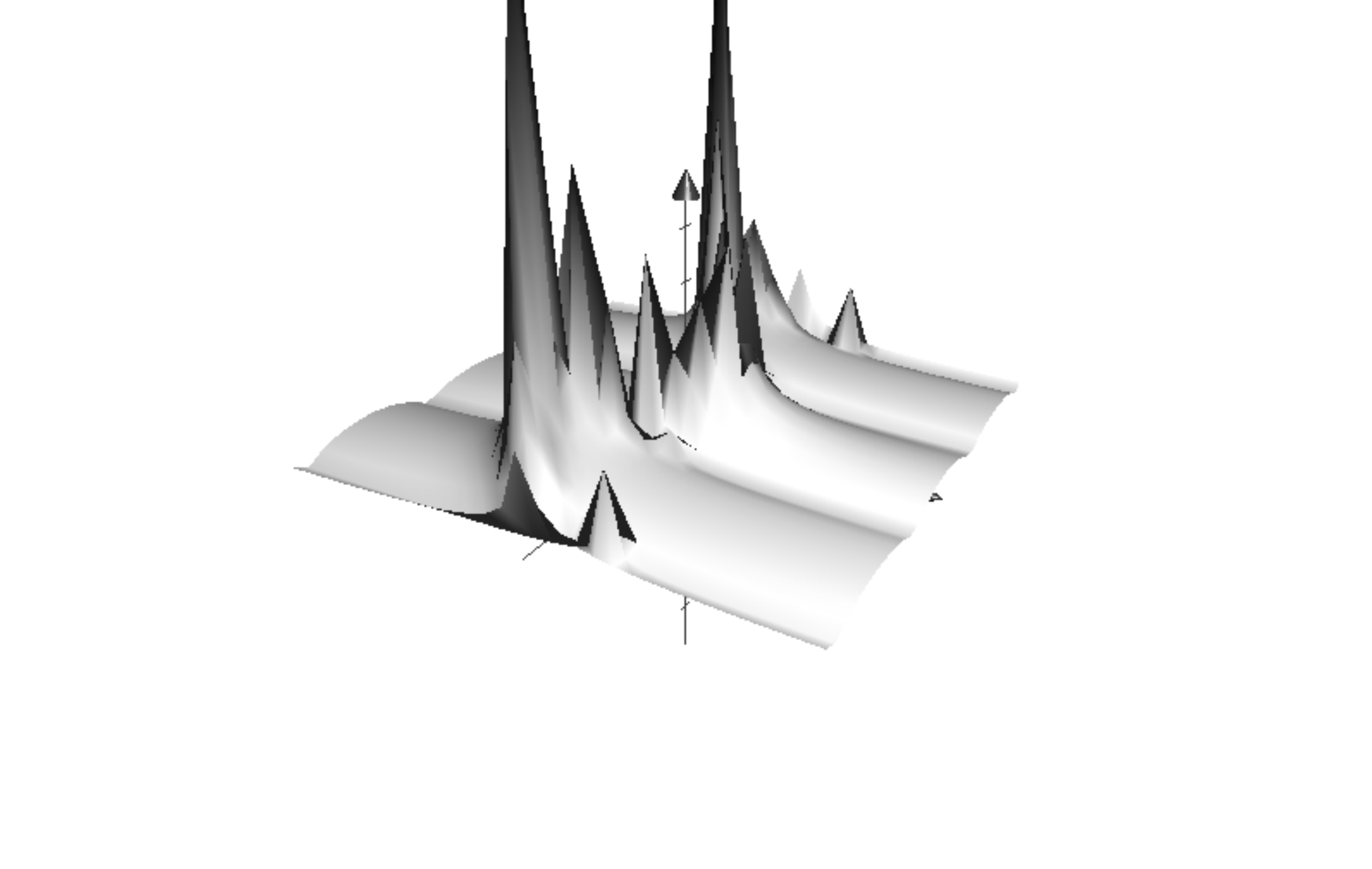}
\caption[The tomogram of a classical de Sitter universe]{The tomogram of the classical de Sitter universe}
\label{fig:classicaltomo}
\end{figure}

\section{The de Sitter quantum cosmological models}\label{desitterquantum}
In quantum cosmology the Hamiltonian constraint  becomes an equation, called the Wheeler De Witt equation, in which the Hamiltonian operator  annihilates  the wave function $ \psi(q) $ 
\begin{equation}\label{quantumhamiltonianconstraint}
	\hat{\mathcal{H}}\psi (q)=0\, .
\end{equation}
The wave function represents the state of the quantum universe.

 The Wheeler-DeWitt equation in the de Sitter universe is
\begin{equation}\label{WheelerDeWitt}
\left(4\hbar^{2}\frac{d^{2}}{dq^{2}}+\lambda q -1 \right)\psi(q) =0\, .
\end{equation}

 The properties of this equation  were discussed  in \cite{Halliwell:1988ik}. With the change of variables
\begin{equation}\label{key}
x=-\left( \frac{\lambda}{4\hbar^{2}}\right)^{1/3} \left( q-\frac{1}{\lambda}\right) =\frac{1-\lambda q}{\left(2\hbar\lambda \right)^{2/3} }
\end{equation}
 equation (\ref{WheelerDeWitt}) becomes  the Airy equation  
\begin{equation}\label{Airyequation}
 \frac{d^{2}\psi(x)}{d x^{2}}-x\psi(x)=0,
\end{equation}
whose solutions are expressed by the integral on the complex plane
\begin{equation}\label{Airy solution}
\psi(x)=\int_{C}\exp\left(x z-\frac{z^{3}}{3} \right) 
\end{equation}
and depend on the choice of the path $ C $ \cite{abramowitz} \cite{Vallee}. 

There are  has two independent solutions.
 The  solutions of interest are those which take real values. These are the  $ \Ai(x) $ defined  by taking as path $ C $ the imaginary axis from $ -i\infty $ to $ +i\infty $ so that its integral representation is 
\begin{equation}\label{key}
\Ai(x)=\dfrac{1}{2\pi}\int_{-\infty}^{+\infty}\exp\left[ {i\left( \dfrac{z^{3}}{3}+xz\right) }\right] dz
\end{equation}
and  $ \Bi(x) $  which is  is obtained taking the path  $ C  $ following  the imaginary axis from $ +i\infty $ to $ 0  $ and then along the real axis from $ 0  $ to $ +\infty $, its  integral representation is
\begin{equation}\label{key}
\Bi(x)=\dfrac{1}{\pi}\int_{0}^{+\infty}\left[ \exp   \left( -\dfrac{z^{3}}{3}+xz\right) +\sin\left( \dfrac{z^{3}}{3}+xz\right) \right] dz\, .
\end{equation}
$ \Bi(x) $ has the property to be real valued for real  $ x $.
Any other solution of the Airy equation is given by a linear combination of these two functions.  

Among all the solutions we  can select the  physically meaningful ones by imposing appropriate initial conditions.  Of particular interest are the conditions which predict  a classical evolution  with the characteristics observed in the present universe.

In this context, among the various initial conditions proposed the best known and most studied are the  Hartle and Hawking ``no-boundary condition'' and  Vilenkin's  ``tunneling of the universe from nothing beginning ''  which was resubmitted in a different form by Linde.

With  the initial no-boundary solution, Hartle and Hawking have intended to restrict the set of solutions of the Wheeler De Witt equation by selecting the set of 4-geometries allowed  such that   there is not a singular  initial boundary differently from  the classical solutions.  This is obtained by `''smoothing  the geometry of the universe off in an imaginary time''.\cite{Wiltshire:1995vk}. For example, whereas a surface with  $ \sqrt{h} $ = 0 (with $ h $ the determinant of the spatial metric) would be singular in a Lorentzian signature metric, this is not necessarily the case if the metric is of Euclidean signature. In other words it is necessary  to  choose the initial
condition ensuring the closure of the four-geometry\cite{Halliwell:1990uy}

Vilenkin represented  with the tunneling of universe  an  initial state of the universe in an analogy with  tunneling effect  in quantum mechanics. 

As in quantum mechanics the solutions   contain ``ingoing'' and ``outgoing'' modes, in quantum cosmology one has to contracting  and expanding modes.  According yo Vilenkin's  proposal
the universe wave function  consists  of those containing only  expanding modes at the
parts of the boundary of superspace corresponding to the classical singular four-geometries. Moreover a regularity condition, that $ \psi $ be everywhere bounded, is also imposed.\cite{Halliwell:1990uy}

  In this paper we consider  the different wave functions In a de Sitter minisuperspace.  We reproduce the results obtained in  \cite{Halliwell:1988ik}  and extend the discussion of these models to the Wigner  functions  (see \cite{Habib:1990hz}\cite{Habib:1990hx}  and \cite{Cordero:2011xa} ) and to the tomographic representation.

\subsection{ Wave Functions}\label{wavefunctions}

In the  de Sitter quantum cosmological models the initial conditions  are linear combinations of the Airy functions. The form  of the wave  functions corresponding to  the initial conditions was  obtained by Halliwell and Louko in \cite{Halliwell:1988ik}.  It particular they found that  the Hartle and Hawking wave function is

\begin{equation}\label{HartleHawkingwavefunction}
\psi_{HH}=A\, \Ai\left( \frac{1-\lambda q}{\left(2\hbar\lambda \right)^{2/3} }\right), 
\end{equation} 

Vilenkon's  tunneling from nothing is  obtained by the complex combination 

\begin{equation}\label{Vilenkinwavefunction}
\psi_{V}\left( \frac{1-\lambda q}{\left(2\hbar\lambda \right)^{2/3} }\right) =\frac{B}{2}\left( \Ai\left( \frac{1-\lambda q}{\left(2\hbar\lambda \right)^{2/3} }\right) +i\, \Bi\left( \frac{1-\lambda q}{\left(2\hbar\lambda \right)^{2/3} }\right) \right) 
 \end{equation}

and finally  Linde's  condition is given by
 \begin{equation}\label{Linde}
 \psi_{L}=-i C\,\cdot \Bi\left( \frac{1-\lambda q}{\left(2\hbar\lambda \right)^{2/3} }\right)
 \end{equation}
 
 Where $ A $,  $ B $ and  $ C $ are normalization constants which can be determined.

 The    properties of these different wave functions can be better highlighted if we consider the following approximate forms at  large values of $ q $ (see \cite{Vallee}),

\begin{equation}\label{approxHH}
\psi_{HH}\approx \frac{(2\lambda\hbar)^{1/6}}{2(\pi^{2}(\lambda q-1))^{1/4}}(\e^{iS}+\e^{-iS})
\end{equation}

\begin{equation}\label{approxv}
 \psi_{V}\approx i\, \frac{(2\lambda\hbar)^{1/6}}{2(\pi^{2}(\lambda q-1))^{1/4}}\e^{-iS}
\end{equation}

\begin{equation}\label{approxL}
 \psi_{L}\approx i\,\frac{(2\lambda\hbar)^{1/6}}{2(\pi^{2}(\lambda q-1))^{1/4}}(\e^{iS}-\e^{-iS})
\end{equation}
where 

\begin{equation}\label{approxS}
 S=\frac{1}{3\lambda\hbar}\left( \lambda q-1\right) ^{3/2}-\frac{\pi}{4}
\end{equation}
Eqs. (\ref{approxHH})-(\ref{approxL}) show  that the Vilenkin solution actually contains only the expanding modes, while the Hartle and Hawking and the Linde solutions contain both contracting and expanding modes. This behavior is most emphasized when one introduces the functions of Wigner and the tomograms, where the superposition of modes shows the presence of positive and negative interferences in the phase space. For this reason it is interesting to discuss the properties of these functions in relation to the initial conditions. In this way it will be possible to evaluate the major and minor probabilities of evolution of the universe. In the following we will summarize the results obtained for the Wigner functions   and their properties.

\subsection{ Wigner Functions}\label{wginerfunctions}
The Wigner function was introduced by Wigner in his paper ``On The Quantum Correction For Thermodynamic Equilibrium''\cite{Wigner:1932eb} after observing that in quantum mechanics there was not a simple expression for probability as in classical statistical mechanics.
   As a   consequence of  the impossibility to measure simiultaneously  the position and the momenta,   Wigner functions may take negative values. Therefore they  cannot be interpreted as   probability functions.

 In quantum cosmology Wigner functions were introduced because, although the formalism in phase space completely describes  all the aspects of  quantum mechanics, the strength of formalism is its ability of dealing more clearly with semiclassical problems  and the transition from quantum into classical systems.  
 
  Firstly because as noted by Wheeler, the wave function is not suitable for implementing a classic limit because it is not localizable and has multiple peaks.
  
  Secondly, it has been hypothesized by some researchers that the presence in a semiclassical model of a correlation between the canonical variables is  a signature of a classical transition
  
  Finally, it can be established that in quantum cosmology there can not be a classical transition without quantum decoherence and  it has been pointed out by various authors that Wigner's functions proves suitable for studying it. 
  
Often quantum decoherence it is identified with  the cancellation of the interference terms typical of the quantum phenomena.
But it the result of a coarse graining of the quantum system.  It can be achieved in different ways. For example if the relevant measures can be done only by specifying the variables $ q^{i}  $ not at all times $ t $, but in a discrete number of times $ t_{i} $ with $ i=1, 2, 3, \dots $, otherwise specifying not all the variables $ q^{i} $ at any one time, but some of them or finally by not specifying the definite values of the  $ q^{i} $, but only range of values  $ \Delta^{i} $ they can assume\cite{Hartle:1989rb}. 
 
On the other side the quantum decoherence manifests itself in open systems through  their interaction with an external environment, similarly in quantum cosmology the same is obtained by implementing a coarse graining obtained through a trace-over unobservables.
  
Some authors tried to find particular solutions of the Wheeler-DeWitt equation peacke araound a subset of solution of the field equation, so that the boundary conditions of the Wheeler-DeWitt equation lead, in the classical limit, to the initial conditions on the classical solutions, see for example\cite{Halliwell:1987eu}.  To show this an approximate solution was used to construct the Wigner function. That this procedure was incorrect was shown subsequently in \cite{Anderson:1990vc}, where it was shown that the Wigner functions obtained with   approximate functions and do not retain the same properties of those obtained with the complete solutions as it happens for example with the harmonic oscillator.

 The exact solutions for a de Sitter quantum cosmological model was obtained by\cite{Habib:1990hz}\cite{Habib:1990hx} using 
  the results obtained by Berry \cite{Berry:1977}.  Which can be obtained also by applying the (\ref{Wigner function}) tp the wave functions obtained by Halliwell and Louko in \cite{Halliwell:1988ik}. Finally in  \cite{Cordero:2011xa} a complete analysis of  the Wigner function from the point of view of the Deformed Quantum Cosmology. It was shown that the previous solutions can be obtained from the Moyal equations(for an overview of the definition of the star product and the Moyal equation see \cite{Curtright:2005}).

 \begin{equation}\label{MOYAL}
 \mathcal{H}\star W(q,p)=0
 \end{equation}
  The Hartle-Hawking Wigner function  can be derived directly  from \ref{MOYAL} or by applying  eq. (\ref{Wigner function} ) to the wave function (\ref{HartleHawkingwavefunction})
It takes the form,
\begin{equation}\label{HHWigner}
W_{HH}(q,p)=\frac{2^{1/3}A^{2}}{\pi(\hbar \lambda)^{1/3}}\Ai\left[\frac{4p^{2}-\lambda q +1}{(\hbar \lambda)^{2/3}} \right] .
\end{equation}
 Only in this case it is possible to obtain explictly this Wigner function. In the other two cases   the  wave  function contains the function $ \Bi(q,p)  $ which is not a limited function, even if formally it is a solution of the Moyal equation. It must be excluded because the Wigner function s to be normal or normalizable (see eq. (\ref{normalwignerfunction})).

\section{The quantum tomograms of the universe and the classical limit}\label{quantumtomogramsandtheclassicallimit}
In this section we will stress the validity and effectiveness of the tomographic approach by discussing the properties of the initial Hartle-Hawking conditions of which we have an analytical expression.

\begin{figure}
\centering
\includegraphics[width=1.2\linewidth]{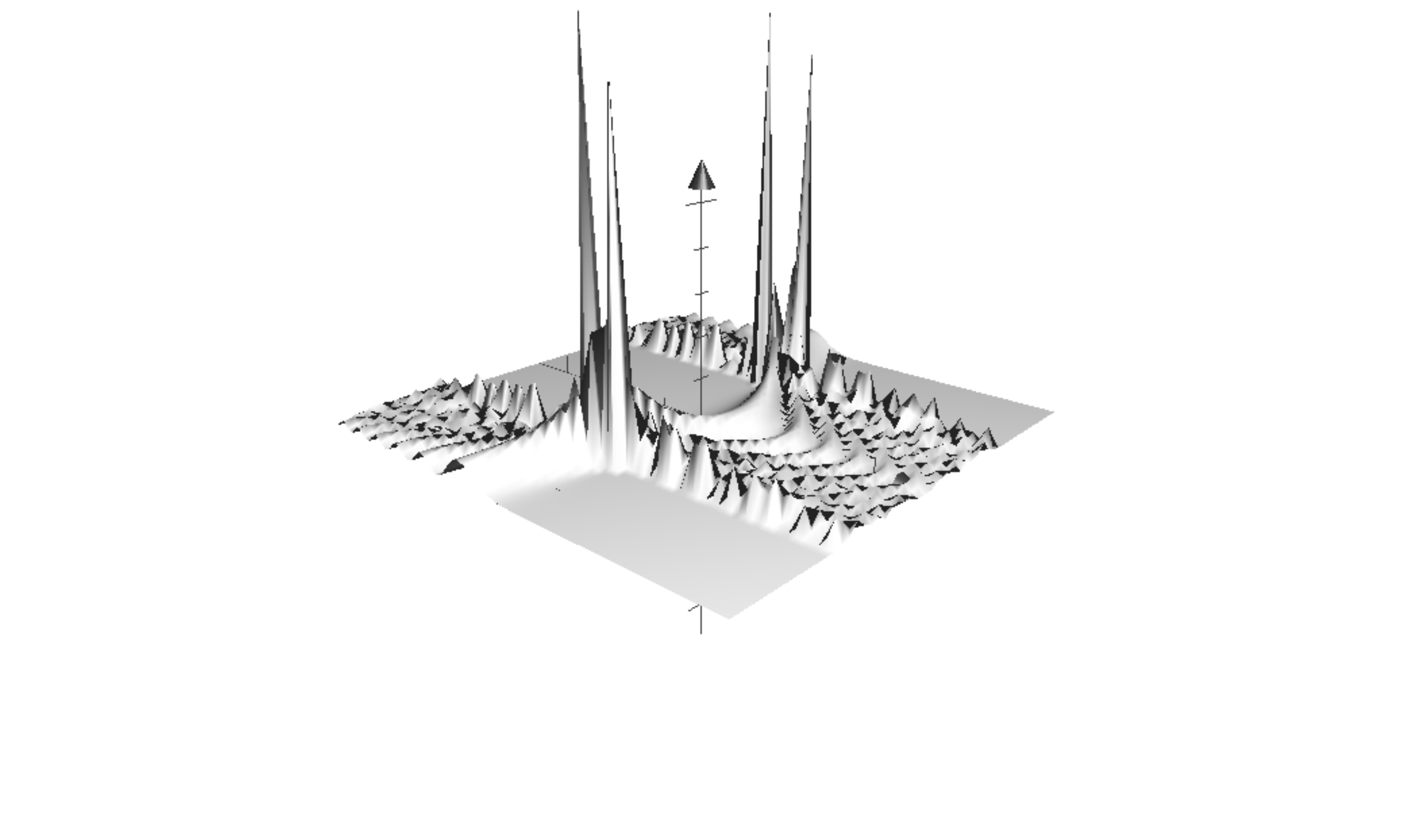}
\caption[The Hartle-Hawking tomogram]{The rich structure of the Hartle Hawking tomogram}
\label{fig:HHtomogram}
\end{figure}

 Let us  derive the tomogram corresponding to the wave function( \ref{HartleHawkingwavefunction}). 
By inserting (\ref{HartleHawkingwavefunction})  in  (\ref{nunonzero})   we immediately obtain 
\begin{equation}\label{tomo_hh}
\mathcal{W}(X,\mu,\nu)= \frac{A^{2}}{|\mu|}\left|\Ai\left( \frac{1}{(2\hbar\lambda)^{2/3}}\left( 1-\frac{\lambda X}{\mu}-\frac{\lambda^{2}}{16}\frac{\nu^{2}}{\mu^{2}}\right) \right)  \right|^{2} 
\end{equation}
where the constant $ A $ will be determined using the asymptotic expression of the tomogram in a similar way to the procedure described  section 21 of \cite{Landau:1991wop} for the one variable wave fucntions.
The wave function is formally the same found for an electrion subject to a linear potential or a constant electric field (see \cite{Landau:1991wop}): In \cite{Manko_Shchukin} the properties of the wave function were discussed  and the corresponding Wigner function and tomogram were obtained.
We illustrate the qualitative  properties of the tomogram in figure, we put $ \lambda=1/2 $ such that $2\hbar\lambda=1 $.  We choose the parametrization  $ \mu=\cos\phi $ and $ \nu=\sin\phi $ (typically use for \textit{optical} tomograms) so that the tomogram could be plotted in the variables $ X $ and $ \phi $.
The graphic of the tomogram is given in figure \ref{fig:HHtomogram}

The determination of the quantum tomogram allows us to study the classical limit.
One of the ways to do this is to determine the limit of the tomogram for $ \hbar\to 0 $ and compare the result with the classical tomogram (\ref{classicaluniversedetomogram}). This problem was addressed in \cite{classicalquantumtransition}.

Let us put 
\begin{equation}\label{hbarwavefunction}
\psi(x)=\hbar^{\gamma/2}\psi(\hbar x)
\end{equation}
so that its norm is invariant. Then we write the tomogram constructed with  (\ref{hbarwavefunction})
\begin{equation}\label{hbartomogram}
\mathcal{W}\left(X,\mu,\nu \right)=\frac{1}{2\pi\hbar|\nu|}\left| \int\hbar^{\gamma/2} \psi\left( \hbar^{\gamma} y\right)  \exp\left[ {i\left( \frac{\mu}{2\hbar\nu}y^{2}-\frac{X}{\hbar\nu}y\right)}\right] \right| ^{2} dy 
\end{equation}
to  calculate the limit $ \hbar\to 0 $ we apply the delta theorem which states that for any
 function $ f(x) $  such that 
\begin{equation}\label{N}
\int_{-\infty}^{\infty}f(x)=N
\end{equation}
  we can show that it has the limit
\begin{equation}\label{deltatheorem}
\lim_{n\to \infty} n f(n(x-x_{0}))=N\delta(x-x_{0}).
\end{equation}
In \cite{Vallee}  this theorem is discussed and demostrated in a different way for the Airy function $ Ai(x) $.

Let us apply this theorem to the Wigner function (\ref{HHWigner}) by considering  the limit $( 2\hbar\lambda)^{2/3}\to 0 $ 
With an appropriate choice of constant A it follows that
\begin{equation}\label{wignerconvergence}
\lim_{ (2 \hbar\lambda)^{2/3}\to 0}W_{HH}(q,p)=\delta(-4p^{2}+\lambda q-1)
\end{equation} 
We   conclude that  from the point of view of the Wigner functions the Hartle and Hawking initial condition has a classical limit.

Let us now examine  the  transition  from quantum to classical from a tomographic point of view.

From (\ref{tomo_hh}) we see that  the limit $( 2\hbar\lambda)^{2/3}\to 0 $ implies that it is sufficient to  use  the asymptotic expressions of the Airy function $ \Ai(x) $. They are
 \begin{equation}\label{approxvallee1}
 \Ai(x)\approx\frac{1}{2\pi^{1/2}\left| x\right|^{1/4} }\e^{-\xi}L(-\xi)
  \end{equation}
  and
  \begin{equation}\label{approxvallee2}
   \Ai(-x)\approx\frac{1}{\pi^{1/2}\left| x\right|^{1/4} }
   \left[  \sin\left( \xi-\frac{\pi}{4}\right) Q(\xi)+\cos\left( \xi-\frac{\pi}{4}\right) P(\xi)\right] 
  \end{equation}
  where  the functions $ L(\xi) $, $ P(\xi) $ and $ Q(\xi) $ are  given by the following series,
\begin{equation}\label{L}\  
L(\xi)=\sum_{s=0}^{\infty }\frac{a_{s}}{\xi^{s}}=1+\frac{3\cdot 5}{1! 216}\frac{1}{\xi}+\frac{5\cdot 7\cdot 9\cdot 11}{2! 216^{2}}\dfrac{1}{\xi^{2}}+....\,
\end{equation}
\begin{equation}\label{P}\  
P(\xi)=\sum_{s=0}^{\infty }(-1)^{s}\frac{a_{2s}}{\xi^{2s}}=1-\frac{5\cdot 7\cdot 9\cdot 11}{2! 216^{2}}\cdot \dfrac{1}{\xi^{2}}+\frac{ 9\cdot 11\cdot 13\cdot 15\cdot 17\cdot 19\cdot 21\cdot23}{4! 216^{4}}\cdot \dfrac{1}{\xi^{4}}+....
\end{equation}
and
\begin{equation}\label{Q}\  
Q(\xi)=\sum_{s=0}^{\infty }(-1)^{s}\frac{a_{2s+1}}{x^{2s+1}}=\frac{3\cdot 5}{1! 216}\frac{1}{\xi}-\frac{ 7\cdot 9\cdot 11\cdot 13\cdot 15\cdot 17}{3! 216^{3}}\dfrac{1}{\xi^{3}}+...\, ,
\end{equation}
where
$ a_{s}=\frac{\Gamma(3s+\frac{1}{2})}{54^{s}s!\Gamma(s+\frac{1}{2})}$
and
$ \xi=\dfrac{2}{3} S^{3/2}$.
In this approximation when the argument is negative the tomogram is
\begin{align}\label{negativeAi}
\nonumber\mathcal{W}(X,\mu,\nu)&\approx   \frac{A^2}{8\pi^{2}\hbar|\mu|}\frac{(2\hbar\lambda)^{4/3}}{\left|1-\frac{\lambda X}{\mu}-\frac{\lambda^{2}}{16}\frac{\nu^{2}}{\mu^{2}}\right|^{1/2}}\\
&\times\left|P(S)\cdot \cos\left( \frac{2}{3}S^{3/2}-\frac{\pi}{4} \right)+Q(S)\cdot\sin \left( \frac{2}{3}S^{3/2}-\frac{\pi}{4} \right)  \right|^{2}§, .
\end{align}
When the argument is positive the approximated tomogram has the form
\begin{equation}\label{positiveAi}
\mathcal{W}(X,\mu,\nu)\approx   \frac{A^2}{8\pi^{2}\hbar|\mu|}\frac{(2\hbar\lambda)^{4/3}}{\left|1-\frac{\lambda X}{\mu}-\frac{\lambda^{2}}{16}\frac{\nu^{2}}{\mu^{2}}\right|^{1/2}}\left| \e^{-S}L(-S)\right|^{2}
\end{equation}
\begin{equation}\label{Sargument}
S=\frac{1}{(2\hbar\lambda)^{2/3}}\left( 1-\frac{\lambda X}{\mu}-\frac{\lambda^{2}}{16}\frac{\nu^{2}}{\mu^{2}}\right) 
\end{equation}
 The normalization constant $A$ can be fixed by requiring that the quantum and classical tomograms have the    same coefficient, i.e.
\begin{equation}\label{Acoefficient}
A=\frac{2^{5/6}\pi}{\hbar^{1/6}\lambda^{2/3}}.
\end{equation} 
Let us determine the limit of the functions (\ref{negativeAi}) and (\ref{positiveAi}) for $( 2\hbar\lambda)^{2/3}\to 0 $ .  First we  notice  that  righthand side of eq.  (\ref{positiveAi})
 is zero because  \[ \lim_{(2\hbar\lambda)^{2/3}\to 0}\e^{-2S}=0. \] 
 There is no contribution to the tomogram from the positive sector of the Airy function.
 \begin{figure} 
  \centering
  \includegraphics[width=0.7\linewidth]{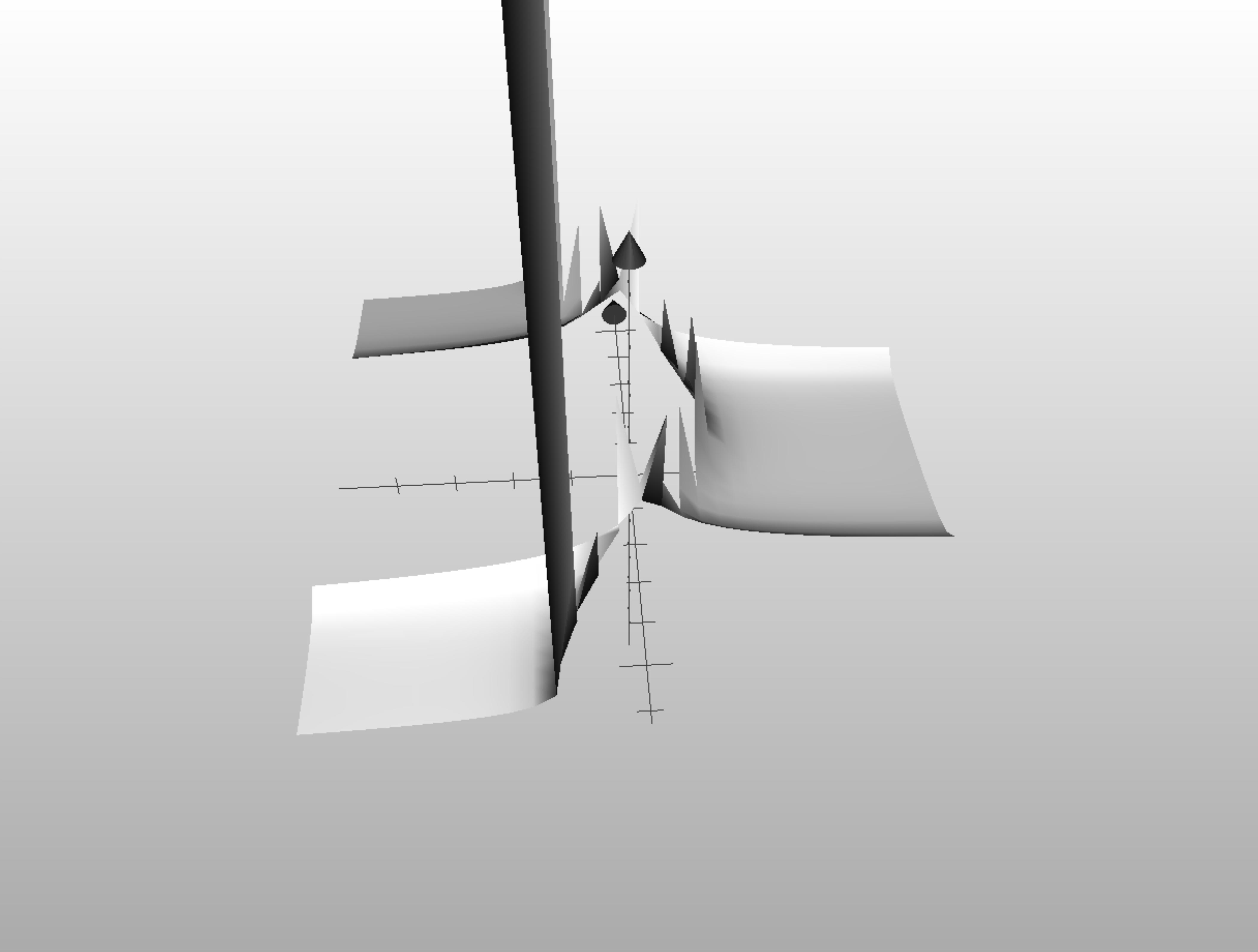}
  \caption{The classical tomogram}
  \label{fig:classicaltomogrammm}
  \end{figure}
  Let us study the negative sector of $ \Ai(S) $. From eqs. (\ref{P}) and (\ref{Q})   we see that 
 $$ \lim_{(2\hbar\lambda)^{2/3}\to 0}Q(S) =0 $$  
 and
 $$ \lim_{(2\hbar\lambda)^{2/3}\to 0}P(S) =1,  $$  
 then in eq.(\ref{negativeAi}) we are left with the square modulus of the cosine, but its argument goes to infinity and the quantum tomogram does not converge to the classical tomogram,  even if one of its factors coincides with the expression (\ref{classicaluniversedetomogram}).  
 
 Even modifying the expression of the tomogram as in (\ref{hbartomogram}) to apply the delta theorem, we do not obtain the classical limit. For example if one adjusts the exponent  $ \gamma $ of $ \hbar $ such that the cosine and the sine go a to a constant, the functions $  P $ and $ Q $ will take both an infinite value. Considering that it is not possible to determine a finite tomogram for the Vilenkin and Linde wave functions, we conclude that none of the proposed initial conditions lead to a classical universe. 
 
The following tomogram
 
\begin{align}\label{possiblesolution}
\mathcal{W}(X,\mu,\nu)&= \frac{A^{2}}{|\mu|}\left|\frac{1}{2}\left( \Ai\left( \frac{1}{(2\hbar\lambda)^{2/3}}\left( 1-\frac{\lambda X}{\mu}-\frac{\lambda^{2}}{16}\frac{\nu^{2}}{\mu^{2}}\right) \right) \right. \right. \\
&\left. \left. +i\Bi\left( \frac{1}{(2\hbar\lambda)^{2/3}}\left( 1-\frac{\lambda X}{\mu}-\frac{\lambda^{2}}{16}\frac{\nu^{2}}{\mu^{2}}\right) \right)\right)  \right|^{2} 
 \end{align}
  witnnh $ A $  given by (\ref{Acoefficient}), provided the  condition (\ref{normalization}) is satisfied is a possbile solution because in the  limit  $( 2\hbar\lambda)^{2/3}\to 0 $  its asymptotic expression  reduces to the classical tomogram .
  
\begin{figure} 
 \centering
 \includegraphics[width=0.7\linewidth]{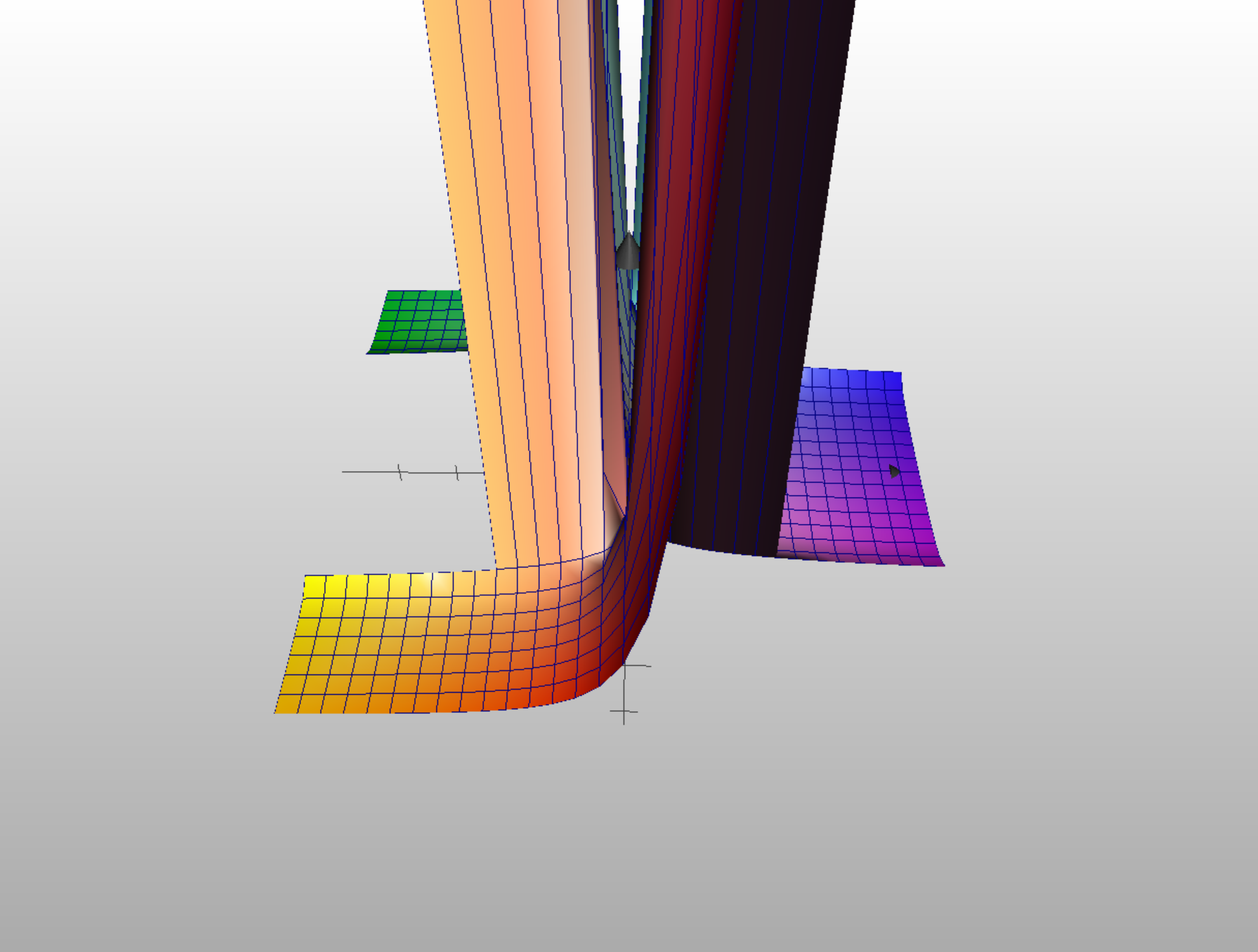}
 \caption{The quantum de Sitter tomogram}
 \label{fig:quantumtomogrammm}
 \end{figure}

\begin{figure} 
 \centering
 \includegraphics[width=1.2\linewidth]{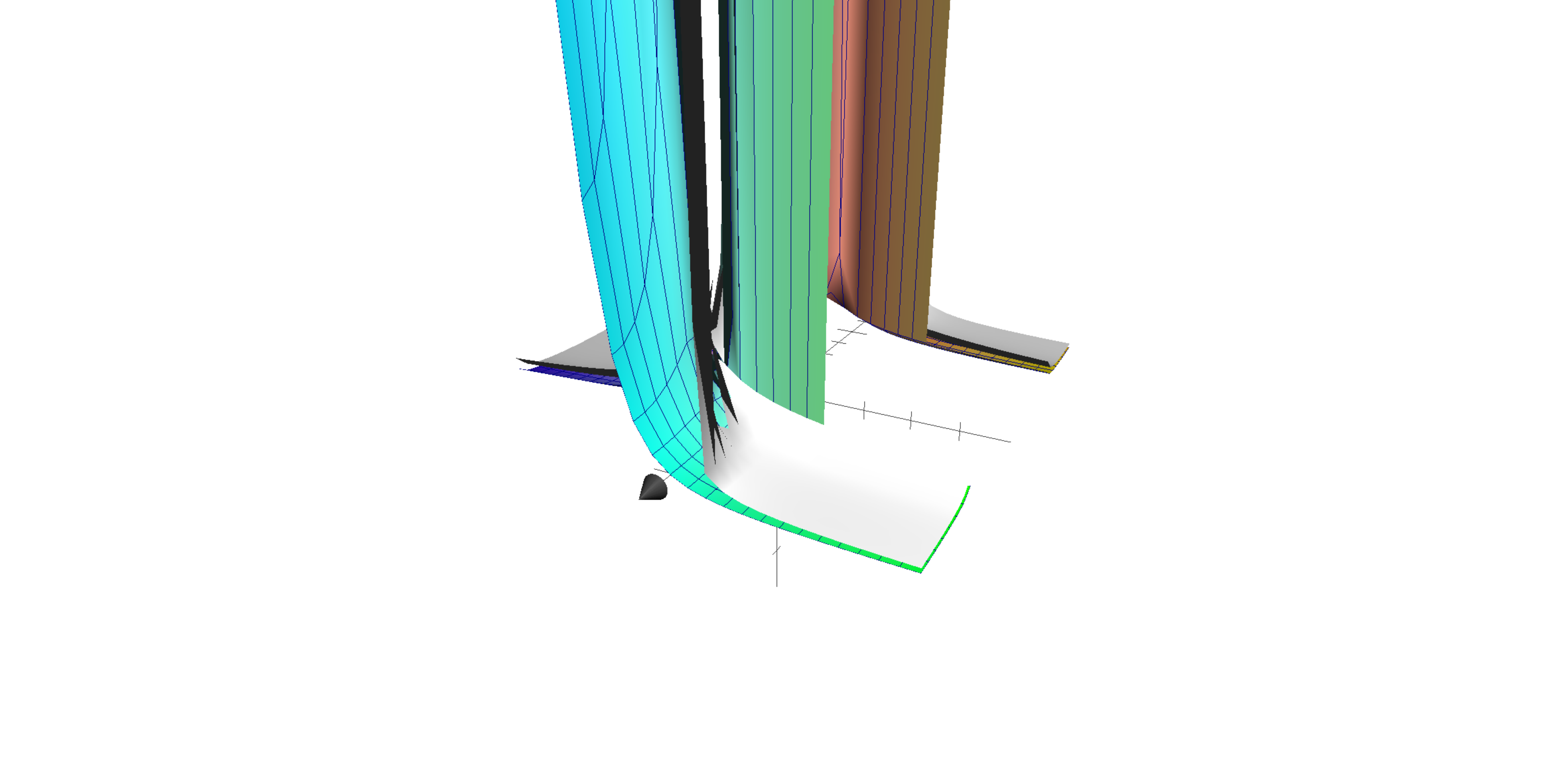}
 \caption{The superposition of the classical and quantum de Sitter tomogram}
 \label{fig:tomogrammaquantumtoclassical}
 \end{figure}

 In this case the classical and quantum tomograms are  compared in figures \ref{fig:classicaltomogrammm}  and \ref{fig:quantumtomogrammm} to show that they can be superimposed see fig.  \ref{fig:tomogrammaquantumtoclassical}.  
  

\section{Perspectives and conclusions}\label{conclusions}
The introduction of the tomographic analysis in quantum cosmology has enabled us   to highlight some important properties of the  solutions of the Wheeler DeWitt equation.

The main result of this paper is the study of the classical limit of a  quantum universe is dominated by the cosmological constant and comparing it with the tomogram derived from the de Sitter classical solution. Although it is an extremely simple model, it still provides us with a very interesting result.

Indeed  as  the classical limit is achieved by taking  the limit  $( 2\hbar\lambda)^{2/3}\to 0 $  in eqs. (\ref{wignerconvergence}) and (\ref{approxvallee1}), we can see   that this limit   is not necessarily obtained  $ \hbar\to 0 $.   Rather it looks like  that the quantum-classical  transition of the universe  depends  on the decay of the cosmological constant from values close to $ \lambda\approx 1 $ to the present value $ \lambda_{c}=5.6\cdot 10^{-122}t^{-2}_{p} $. 
Moreover it results that  the state of the classical universe is constrained by the interval (\ref{compact})  yielding a very high degree of homogeneity.

Therefore the problem of the value of the cosmological constant \cite{Weinberg:1988cp}  becomes crucial to understand the current large-scale structure of the universe.  
We noticed  that even if the  the Hartle-Hawking Wigner function converges to the classical solution, its corresponding  tomogram does not converge because when $( 2\hbar\lambda)^{2/3}\to 0 $ its  asymptotic expression oscillates rapidly , therefore it has not a limit. 

However, we can not rule out that the universe can be described by this particular model. Since the final value of the cosmological constant, although small is different from zero, it could be that the limit of the Hartle and Hawking model describes equally well a universe dominated by the cosmological constant in which the material sources are negligible and where the galaxies can behave as test bodies. Since the integration interval should be of the same order of magnitude as the classical model, the tomogram would present some oscillations that would distinguish it from the classical model (\ref{classicaluniversedetomogram}). An analysis of the distribution of motion of galaxies could falsify one of the two models or even both.

The second interesting point is that the normalization condition (\ref{normalization condition tomogram}) restricts the range  of values that $ X $ can take. This  suggests that  a further investigation is needed to determine  domain of the functions  (\ref{Vilenkinwavefunction})-(\ref{Linde}) such that the resulting tomogram can be expressed, even if in an  analytic way.

On the other side according to the properties of the Airy functions, we introduced the  tomogram  (\ref{possiblesolution}) as a possible solution that converges to the classical solution (\ref{classicaluniversedetomogram}). But    the asymptotic expression of   the oscillating part of the Airy functions  converges  to the inverse square root of  $ |-\xi| $,  and  the  classical solution is  inverse of the modulus of the square root of  $\xi$.  These two functions are not equal everywhere.  The  two tomograms can be  superimposed only  when 
$$
\xi=\frac{\lambda^{2}\nu^{2}}{16\mu^{2}}+ \frac{\lambda X}{\mu}-1\ \ge 0
$$
where the physical relevant regions of definition for $ X $ can be found.

Of course the results of this paper cannot be considered conclusive,  it will be necessary to study more general models taking in account the presence of  scalar fields and  more general material sources.  The method introduced in this paper can be extended to many models in the vast literature of quantum cosmology from the traditional solutions of the Wheeler De Witt equation to the  loop quantum cosmology \cite{Bojowald:2008zzb} and for the further developments (see for example \cite{Bojowald:2015iga}).

Finally the tomographic approach can be extended beyond the  analysis of the single wave functions of the universe,  because the task of quantum tomography is the reconstruction of the initial state of a physical system starting from the experimental data, so with this perspective further work will be devoted to the phenomenological construction of the tomograms of the classical universe.

\end{document}